\documentclass{article}
\pdfoutput=1
\usepackage{geometry}
\usepackage{subfigure,stackrel,multirow}
\usepackage{amsmath,amsfonts,amssymb,amsthm}
\usepackage{bm}
\usepackage{graphicx, tabularx, array}
\usepackage{subfigure} 
\usepackage{color}
\usepackage{booktabs}

\title{A Novel Chronic Disease Policy Model}
\author{Nathan Green, Duncan Smith, Matthew Sperrin, Iain Buchan}

\begin{document}
\maketitle
\begin{abstract}
We develop a simulation tool to support policy-decisions about healthcare for chronic diseases in defined populations. Incident disease-cases are generated in-silico from an age-sex characterised general population using standard epidemiological approaches. A novel disease-treatment model then simulates continuous life courses for each patient using discrete event simulation. Ideally, the discrete event simulation model would be inferred from complete longitudinal healthcare data via a likelihood or Bayesian approach. Such data is seldom available for relevant populations, therefore an innovative approach to evidence synthesis is required.  We propose a novel entropy-based approach to fit survival densities.  This method provides a fully flexible way to incorporate the available information, which can be derived from arbitrary sources.  Discrete event simulation then takes place on the fitted model using a competing hazards framework.  The output is then used to help evaluate the potential impacts of policy options for a given population.
\end{abstract}
\footnotetext{\emph{Address for correspondence}: Nathan Green, North West Institute for Bio-Health Informatics, University of Manchester, Oxford Road, Manchester, M13 9PL.\\
\textsf{Email: nathan.green@manchester.ac.uk \normalfont}}

\section{Introduction}
\label{chapter:intro}
Common chronic diseases, such as cardiovascular diseases, are a major public health problem and consume large amounts of scarce healthcare resources \cite{BHF_stats}.  In order to use limited healthcare resources to best effect for local populations, there is a need for decision-support tools when considering different medical and public health approaches to chronic diseases.  This is a complex area beyond the information resources of most healthcare commissioning organisations.  So we consider how commissioners might explore the potential impacts of different policy options via a flexible and accessible model.
In particular, we shall present a coronary heart disease (CHD) simulation model that is both flexible and useful to commissioners and providers.
We build a discrete event simulation model that accounts for the population evolution over time to better understand the process behind CHD.
Ideally, the model would be inferred from complete longitudinal data for a relevant population via a likelihood or Bayesian approach.
Unfortunately, such data is not readily available and innovative ways are needed to combine information from different sources.  
Therefore, a graph representation and survival analysis methods have been combined with a novel application of simulated annealing in order to learn the temporally dependent distributions of the model.

Our work incorporates and extends several aspects of previous chronic disease modelling approaches.  
The IMPACT I model \cite{Erdos01} was the case study for the model described in this paper. It is a population-level, spreadsheet-based model using simple epidemiological and economics methods to estimate the number of deaths prevented or postponed (DPP).  The model is theoretically simple, and has undergone extensive validation, highlighting limitations that gave rise to the work presented here.
Similar to our model, the Weinstein model, \cite{weinstein1987}, is divided into a population health (primary-side) and a patient disease module (secondary-side). The Weinstein model also includes a \emph{bridge} module which controls an individual's fate in the first 30 days after developing CHD. Despite being a relatively old, discrete time model, this is one of the more sophisticated in this field.
Similar to the Weinstein model, the CHD Life Expectancy Model considers individuals in discrete time. It is, however, less sophisticated that the Weinstein model in the sense that there are fewer states and the time periods are fixed.
Closest to our modelling approach is the CHD Policy Analysis Model \cite{babad02}. Again, it is divided in to a primary side and a secondary side, with model parameter estimates derived directly from the literature.
The Archimedes model is a micro-simulation model, which simulates patients to a high level of detail by focusing on the underlying pathophysiological processes that determine disease \cite{archimedes2002}. The model has predicted the outcomes of notable clinical trials well, and has significant potential generalisability. Archimedes is closed proprietary work, therefore it is not possible to examine the details of the model implementation.  This is concerning given the model complexity and the public health interest involved.
An epidemiological review of CHD policy models is given in \cite{Unal2006}.

This document is structured as follows.
In Section~\ref{section:background}, the mathematical preliminaries are given.
Then, in Section~\ref{section:model_description} we describe how these ideas are used to build the general policy model.
Section~\ref{section:baseline} then specifies a baseline scenario and the context of the paucity of relevant data available.
In Section~\ref{section:fitness}, we explain how the model is fit in light of the data.
Section~\ref{section:interventions} investigates how the model is modified under certain interventions.
Section~\ref{section:example} applies the approach to a specific motivating CHD example. 
Finally, Sections~\ref{section:discussion} and \ref{section:conclusions} summarise and draw conclusions with ideas for future work.

\section{Background}
\label{section:background}
In this section, we introduce existing theory and define the notation necessary for the subsequent sections.

Letting $T$ denote event time, a hazard function is defined as a limiting probability,
\begin{equation}
\label{eqn:general_hazard}
h(t) = \lim_{\delta\downarrow 0} \mbox{P}(t \leq T \leq t + \delta \vert T \geq t)/\delta.
\end{equation}
Further, consider collections of \emph{competing} hazards or causes of failure, indexed by integers 1 to $K$ and with corresponding latent failure times $T_1, T_2, \ldots, T_K$.
The \emph{marginal hazard} is defined as the hazard of event $k = 1, 2, \ldots, K$ occurring, ignoring the possibility that another event $k' \neq k$ may have occurred first.
Suppose that each event has an associated marginal hazard function given by $h_k(\cdot)$, $ k=1,\ldots,K$. Calculation of the overall hazard assumes independence between the competing hazards. Specifically, this is, `the time of failure from cause $k$ under one set of study conditions in which all $K$ causes are operative is precisely the same as under an altered set of conditions in which all causes except the $k^{\textrm{th}}$ have been removed' \cite{prentice78-cr}.

From (\ref{eqn:general_hazard}), the marginal hazards are defined as
$$
h_k(t) = \lim_{\delta\downarrow 0} \mbox{P}(t \leq T_k \leq t + \delta \vert T_k \geq t)/\delta, \;\; k=1,\ldots,K.
$$
The first failure occurs at time $T = \textrm{min}_k \{T_k; k=1,\ldots,K\}$.
Let the corresponding cause of failure at time $T$ be denoted by $C = \textrm{argmin}_k \{T_k; k=1,\ldots,K\}$.
Then, we are interested in the joint distribution of $(T,C)$.

Note that we do not observe the full set of failure times $T_1, T_2, \ldots, T_K$, we only observe the one actual failure time $T$. Therefore, if we had real data, it would not be possible to estimate the model \cite{prentice78-cr}. In particular, it is impossible to identify the dependence structure from data. On the other hand, any collection of competing hazards with any dependence structure can be generated from a related collection of independent competing hazards \cite{Cox59}.

A \emph{sub-hazard} distribution (also known in the literature as a cause-specific hazard) will be of central importance. The sub-hazard distribution describes the hazard of failing specifically from a given cause (rather than any of the competing causes).
From~(\ref{eqn:general_hazard}), this is defined as
$$
h^{\textrm{sub}}_k(t) = \lim_{\delta\downarrow 0} \mbox{P}(t \leq T_k \leq t + \delta \vert T \geq t)/\delta, \;\; k=1,\ldots,K.
$$
Finally, under independence of competing hazards, the \emph{overall hazard} is given by
$$
h(t) = \sum_{k=1}^K h^{\textrm{sub}}_k(t).
$$
The difference between the sub-hazard and the marginal hazard is that the marginal hazard describes the failure from a particular cause \emph{ignoring any other causes}, in contrast to the sub-hazard which describes the failure from a particular cause \emph{rather than any other cause}. Under independent competing hazards, though, these are equal in distribution, $h^{\textrm{sub}}_k = h_k$.

In the interest of brevity, for the remainder of this paper, the ``sub" superscripts will be suppressed and we will assume reference to the sub-distributions, sub-densities and sub-hazards unless otherwise stated.

\section{Model description}
\label{section:model_description}
The model is constructed as a graph consisting of edges $e \in E$ and vertices $v \in V$, which represents possible states and transitions between states. We shall fix the vertices (states) and edges (possible transitions) in advance.  
The graph is used as a framework for a discrete event simulator such that for each individual a sequence of events, from a possible set of events defined by the graph, occur chronologically \cite{DES}.
The topology of the graph determines that competing risks exist. Further, this can be thought of as a multi-state model, an extension of the standard competing risks model \cite{CompetingRisks_Tutorial}.

Revisiting a previous state is possible, so loops in the graph may exist.
A subset of states are specified as entry states, where an individual has presented with CHD symptoms, and another (possibly overlapping) subset of states are defined as sink states, corresponding here to death events. 
A continuous time multi-state model is developed here, which allows higher fidelity simulation and more flexible, realistic intervention policies than a discrete time analogue.

\subsection{Form of the sub-hazard functions}
The hazard functions used in this model are combinations (or mixtures) of $L$ component hazard functions,
\begin{equation}
\label{eqn:general_mixture}
h_k(t) = \sum_{i=1}^L  h_{k, i}(t), \;\; k=1,\ldots,K.
\end{equation}
Using a mixture hazard function enables us to better fit the model to data. Different individual hazard functions may have contrasting properties, so when combined together provide a flexible distribution.
Note that we can even think of the combinations of hazards as competing hazards themselves.
The choice of the number of components in the mixture is part of the model fitting process.

We shall assume that the mixture hazard function has the proportional hazards property; that is, multiplying the hazard function by a constant leads to a hazard function of the same form. This will be exploited when interventions are included in the model.

\section{Baseline model}
\label{section:baseline}
Subsequently, we shall investigate different intervention strategies, but first we produce a baseline model which describes the current progression or trajectory of each individual.
That is, we wish to describe the patient journeys under existing healthcare protocols.
Ideally, we would want to specify a natural progression baseline model where no healthcare interventions are made. However, this data is not available since whenever an individual presents with CHD symptoms treatment is offered and rarely refused.

To determine the model, approximations to the true underlying functions have been produced from a range of data types from various different sources, and expert elicitation.
These approximations are specified through edge and node \emph{constraints} which are used in the model fitting process.
In principle, these constraints may be specified on any of the (sub-) distribution, density, hazard or survival functions.
We thus define a generic constraint function $\tilde{\phi}^j(t_j,\Delta_j)$ to include all of these, where the total number of constraints is $N$, the constraint index is $j=1,2, \ldots,N$, $t_j$ is the initial time and $\Delta_j$ is the time period.

In practice, the majority of the constraints are specified on the sub-distribution functions. 
In this case, we assume that these approximate the transition probabilities, usually with time period of one year. The transition probabilities essentially constitute a discrete-time Markov jump process.
It is only acceptable to treat transition probabilities as sub-distribution functions if a reasonable amount of time is spent in each state. To this end, one-year transition probabilities are used in states that patients are expected to spend a considerable amount of time in, whilst shorter time periods are used in acute states.

The sub-distribution constraints are usually generated at approximately ten year age intervals. For example, a constraint may be specified on the transition from state $x$ to $y$ as $\mbox{P}(y \mbox{ at age } 51 \; | \; x \mbox{ at age } 50)$, with another constraint at $\mbox{P}(y \mbox{ at age } 61 \; | \; x \mbox{ at age } 60)$, and so forth.
We shall also assign different importances to each constraint by allocating a weight $w_i$, $i=1,\ldots,N$. This weight is designed to reflect the amount of data, or strength of opinion, upon which a given constraint is based.

\section{Model fitting}
\label{section:fitness}
This section details how the constraints, derived from expert elicitation and data, are used to approximate underlying continuous functions. This is achieved through a novel application of simulated annealing and an established entropy measure derived from information theory.

Constraints can be thought of as a list of conditions that we wish the model to satisfy.
Define the generic fitted functions $\widehat{\phi}(t)$ to be continuous functions on $t$ which assume some parametric form according to the modelling assumptions.
For example, the fitted sub-hazard functions are of the form given in (\ref{eqn:general_mixture}) and are defined on each edge.
A constraint is satisfied exactly when $\widehat{\phi}^j(t_j,\Delta_j)=\tilde{\phi}^j(t_j,\Delta_j)$, $j=1,\ldots, N$.
The model fitting procedure is a trade-off between satisfying a series of constraints, and conforming to a desired model structure. This can be thought of as a compromise between fit (through the constraints) and avoiding excessive complexity (through specifying a simple model structure).
For flexibility, there are separate models for males and females, which are fitted separately.

Suppose the fitted values for the constraints computed from the model $M$ are $\widehat{\phi}^j(t_j,\Delta_j)(M)$, $j=1,\ldots, N$. The dependence on the model $M$ is explicit here --- if we change the model, the fitted values change also.
We are required to specify a model space that we are allowed to work over; call this model space $\mathcal{M}$, by which we mean
\begin{eqnarray*}
\mathcal{M} & =	& \{\textrm{Models in which all edges have a given mixture hazard function}\\
						& 	& \;\; \textrm{and conform to a fixed graph }(V,E) \}.
\end{eqnarray*}
We call a specific model $M$, so each $M \in \mathcal{M}$.
The reason that the constraints cannot usually be satisfied exactly is that the chosen mixture distribution usually has only four free parameters, and each mixture needs to satisfy more than four constraints.

In the model fitting procedure, we want to find the optimal model $M^*$, such that the difference between the supplied and fitted constraints are minimised, i.e.
$$
M^* = \stackrel[M \in \mathcal{M}]{}{\textrm{argmin}} \sum_{j=1}^N w_j d \left\{ \tilde{\phi}^j(t_j,\Delta_j), \widehat{\phi}^j(t_j,\Delta_j)(M) \right\},
$$
where $d$ is chosen as the Jensen-Shannon distance measure. 
A simulated annealing algorithm is used to minimise the given criterion function. The fundamental idea is to change the scale, or temperature, to allow different moves on the surface of the objective function, thus avoiding being trapped in local minima \cite{Robert_MCStatsMethods}. A detailed exposition of this application will appear in a companion paper.

A step-by-step algorithm description is given in the Appendix.

\section{Interventions}
\label{section:interventions}
An important reason for developing the model was to support the investigation of different intervention strategies in policy scenarios which cannot be feasibly tested in the real world, due to population, financial, or ethical reasons. The model interventions affect patients indirectly, meaning that the results of a simulated intervention do not immediately alter the state of an individual 
but rather alter the probabilities of entering a given state.
That is, an intervention, such as a change in medical treatment, is designed to alter the disease trajectory of a patient.

The incorporation of an intervention is transparent.
A given intervention has proportional hazards adjustments associated with it for transitions from one state to another. It is an assumption of the model that the adjustment in the hazard should be proportional. These are incorporated into the model by simply scaling up or down the marginal hazard function by the relevant amount. Define this relative risk reduction by $\eta$. Each intervention has a given uptake level $p$, which may depend on age and gender. Upon entry to a state, an individual receives a given intervention with probability $p$.

For example, suppose that an individual enters state Acute Myocardial Infarction, \texttt{AMI}. Immediately, we decide whether any of the possible interventions will be applied. Suppose there is only one intervention to consider, `in hospital CPR'.
Assume that we have \emph{a priori} specified the proportion of the population that has this intervention in this state as $p$. If the intervention is applied, let the auxiliary variable $r=1$, otherwise $r=0$.
Thus, sample $u \sim \mbox{Unif}[0,1]$ and if $u < p$ then set $r=1$ and apply the intervention, otherwise do not apply the intervention and set $r=0$.
If we intervene, then suppose that the specified hazard adjustment factor for a transition \texttt{AMI} $\rightarrow$ \texttt{CHD Death} is $\eta=2/3$. Thus, we would have
\begin{equation}
\label{twostep_adjustment}
h_k^{\textrm{new}}(t) = \left( \frac{2}{3} \right)^r h_k^{\textrm{old}}(t),
\end{equation}
for that particular marginal hazard function.

The proportional hazard adjustment depending on a Bernoulli random sample can be thought of in terms of a frailty model~\cite{Collett2003}, a natural extension to the basic Cox model.
The notion of a frailty model is a convenient way to introduce random effects in to the model. Simply, frailty is an unobserved random proportionality factor that modifies the hazard function of an individual. 
Thus, it is a way of transferring the random selection directly on to the hazard. We do not assume that the population is homogeneous but rather a heterogeneous sample, where different individuals have different hazards, in our case either adherent or not. 

The univariate frailty model is such that the hazard of an individual depends in addition on an unobservable random variable $Y$, which acts multiplicatively on the baseline hazard function. Analogous to (\ref{twostep_adjustment}),
\begin{displaymath}
h_k^{\textrm{new}}(t) = Y h_k^{\textrm{old}}(t).
\end{displaymath}
In particular, we have the simple case where the relative risk adjustment only takes one of two values, so setting $Z \sim$Bernoulli($p$) we obtain 
\begin{displaymath}
h_k^{\textrm{new}}(t) = \eta^Z h_k^{\textrm{old}}(t).
\end{displaymath}
The expect adjusted hazard $\mathbb{E}[h_k^{\textrm{new}}(t)] = \{1+p(\eta-1)\} h_k^{\textrm{old}}(t)$. This can be thought of as the associated proportional adjustment combining the uptake probability and the relative risk. When the variance $\mbox{Var}[h_k^{\textrm{new}}(t)] = \{(\eta - 1) h_k^{\textrm{old}}(t)\}^2 p(1-p)$, is small or the population size is large then using $\mathbb{E}[h_k^{\textrm{new}}(t)]$ may be more practical whilst not adversely affecting accuracy.

\section{Example}
\label{section:example}
In this section, we will give a CHD example implementation with interventions.

\subsection{CHD model description}
\begin{figure}
	\centering
		\includegraphics[width=10cm]{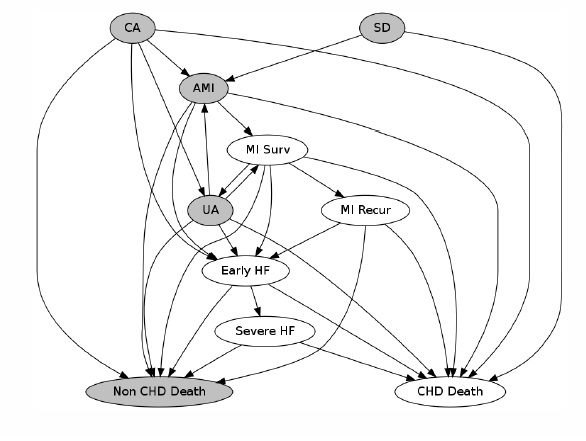}
				\caption{Graphical model of the secondary side. This is a multi-state model, where the nodes represent states that patients can belong to, and the edges represent possible transitions between the nodes.}
	\label{fig:secondary}
\end{figure}

Figure~\ref{fig:secondary} shows the graph structure and Table \ref{tab:NodeNamesTable} gives the node descriptions from this graph.
For ease of exposition, we shall assume that both males and females have the same graphical structure.
The graph can be described as follows:
Both death states (\texttt{Non CHD Death} and \texttt{CHD Death}) can be caused by any non-death state.
Early Heart Failure (\texttt{Early HF}) can be caused by Unstable Angina (\texttt{UA}), Chronic Angina (\texttt{CA}) or various stages of Myocardial Infarction (\texttt{AMI}, \texttt{MI Surv}, \texttt{MI Recur}).
Unstable Angina can recur after Acute Myocardial Infarction, which itself can result from either Chronic Angina or Sudden Death (\texttt{SD}), and so forth.

\begin{table}
\caption{\label{tab:NodeNamesTable} CVD model node description}
		\begin{flushleft}
		\begin{tabular}{|lll|}
			\toprule
			Node Name 					& Full Name 						& Description \\
			\midrule
			\texttt{CA} 				& Chronic Angina 				& long term severe chest pain\\
			\texttt{AMI} 				& Acute Myocardial Infarction & heart attack\\
			\texttt{SD} 				& Sudden Death 					& heart stopping\\
			\texttt{MI Surv} 		& Myocardial Infarction Survival & recovered heart attack\\
			\texttt{MI Recur}		& Myocardial Infarction Recurrence & recurrence of a heart attack\\
			\texttt{UA} 				& Unstable Angina 			& triggered severe chest pain\\
			\texttt{Early HF} 	& Early Heart Failure 	& weak heart\\
			\texttt{Severe HF} 	& Severe Heart Failure 	& very weak heart\\
			\texttt{Non CHD Death} & Non CHD Death 			& death by a cause other than CHD\\
			\texttt{CHD Death}	& CHD Death 						& death caused directly by CHD\\
			\bottomrule
		\end{tabular}
			\end{flushleft}
\end{table}

The baseline model considers a scenario where current guidelines are followed. The interventions and the related nodes on which they act are given in Table~\ref{table:interventions_baseline}.

\begin{table}
\caption{\label{table:interventions_baseline} Baseline Interventions}
		\begin{tabular}{|ll|}
			\toprule
			Intervention Name						& States	\\
			\midrule
			ACE inhibitors							&	\texttt{CA}, \texttt{AMI}, \texttt{SD}, \texttt{MI Surv}, \texttt{MI Recur}, \\
																	& \texttt{UA}, \texttt{Early HF}, \texttt{Severe HF} \\	
			Aspirin											&	\texttt{CA}, \texttt{AMI}, \texttt{SD}, \texttt{MI Surv}, \texttt{MI Recur},\\
																	& \texttt{UA}, \texttt{Early HF}, \texttt{Severe HF}\\
			Aspirin and low molecular weight heparins							&	\texttt{UA}								\\
 			Glycoprotein IIb/IIIa inhibitors &	\texttt{UA}		\\
 			Warfarin										&			\texttt{CA}, \texttt{MI Surv}, \texttt{MI Recur}				\\ 																										 
			Beta blockers in heart failure					&	\texttt{Early HF}, \texttt{Severe HF}								\\
			Beta blockers in SP					&	\texttt{CA}, \texttt{MI Surv}, \texttt{MI Recur}						\\
			Early beta blockers					&	\texttt{AMI}						\\
 			In-hospital resuscitation							&			\texttt{AMI}					\\
 			Out-of-hospital resuscitation					&		\texttt{SD}					\\
 			Percutaneous coronary intervention (PCI) CA&	\texttt{CA}\\
 			Primary PCI									&			\texttt{AMI}						\\
 			ST-elevation myocardial infarction (STEMI) PCI&	\texttt{UA} \\  
 			Coronary artery bypass graft (CABG)&	\texttt{CA}							\\
			CABG and non-STEMI									&		\texttt{UA}							\\
 			Primary CABG								&		\texttt{AMI}						\\
 			Rehabilitation							&			\texttt{CA}, \texttt{MI Surv}, \texttt{MI Recur}, \texttt{Early HF},\\
 																	&			 \texttt{Severe HF}			\\
 			Spironolactone							&			\texttt{Early HF}, \texttt{Severe HF}			\\
 			Thrombolytics								&			\texttt{AMI}				\\
 			Statins											&			\texttt{CA}, \texttt{MI Surv}, \texttt{MI Recur}, \texttt{Early HF},\\
 																	&		 \texttt{Severe HF}			\\
 			\bottomrule
		\end{tabular}
\end{table}

The percutaneous coronary intervention (PCI), commonly known as coronary angioplasty or stenting, is a non-surgical way to widen obstructed blood vessels in the heart muscle.   
Coronary artery bypass graft (CABG), is surgery that takes arteries or veins from elsewhere in the patient's body and uses them to bypass the narrowed blood vessels supplying the heart muscle.
Angiotensin-converting enzyme (ACE) inhibitors are drugs that relax blood vessels and help to decrease blood pressure.
Spironolactone is a drug that gets rid of excess fluid without lowering potassium levels too much, and helps to reduce blood pressure.
Thrombolytics are drugs given just after a heart attack to dissolve clots in coronary arteries.
Statins are drugs that lower cholesterol and help to slow down the furring up of arteries with fats.
Warfarin, heparins and gylyoprotein inhibitors are drugs that inhibit blood clotting.
Beta blockers are drugs that reduce the work of the heart thereby reducing the risk of angina.
Myocardial infarction is heart attack.
Rehabilitation is exercise for the heart muscle to build it up again after a heart attack.
Rescuscitation is restarting the heart and lungs after they have stopped.

\subsection{Inputs}
We will investigate the effect on a population of size 10,000 of different levels of preventative treatments regimes. The population age structure, gender distribution and starting states are derived from the well-known ASSIGN cohort \cite{ASSIGN}.
We shall consider 3 separate scenarios: i) no interventions; ii) beta blockers only iii) statins, beta blockers, aspirin and ACE inhibitors. The scenarios have been chosen for simplicity of exposition.

Table~\ref{table:uptake} gives the uptake values on each node for the considered treatments.
Tables~\ref{table:RR_ACEI}, \ref{table:RR_Aspirin}, \ref{table:RR_BetaBlockers} and \ref{table:RR_Statins} give the relative risk reductions on each edge for ACE inhibitors, aspirin, beta blockers and statins respectively.

\begin{table}
\caption{\label{table:uptake} Uptake of interventions (\%)}
		\begin{tabular}{|lllll|}
			\toprule
			State Name					& ACE Inhibitors	& Aspirin	& Beta Blockers	& Statins \\
			\midrule
	 		\texttt{AMI} 				& 18 & 93 & 4  &\\
			\texttt{SD} 				&    & 100&    &\\
			\texttt{CA}					& 76 & 93 & 77 & 88\\
			\texttt{MI Surv} 		& 76 & 93 & 77 & 88\\
			\texttt{MI Recur}		& 76 & 93 & 100& 86\\
			\texttt{UA} 				&    & 93 &    &\\
			\texttt{Early HF} 	& 80 & 82 & 62 & 82\\
			\texttt{Severe HF} 	& 80 & 82 & 62 & 82\\
			\bottomrule
		\end{tabular}
\end{table}

\begin{table}
\caption{\label{table:RR_ACEI} Risk reductions for ACEI (\%)}
		\begin{tabular}{|llllllllll|}
			\toprule
								&	\!\texttt{AMI}\!\!	& \!\texttt{SD}\!\!	& \!\texttt{CA}\!\!	& \!\texttt{MI Surv}\!\!	& \!\texttt{MI Recur}\!\!	& \!\texttt{UA}\!\!	& \!\texttt{Early HF}\!\!	& \!\texttt{Severe HF}\!\!	& \!\texttt{CHD Death}\!\!\\
			\midrule
	 		\!\texttt{AMI}\!&								&										&										&						93						&													&										&				93								&														& 79\\
			\!\texttt{SD}\!&								&										&										&													&													&								 		&													&														&   \\
			\!\texttt{CA}\!&		93					&										&										&													&													&								 93	&				93								&														& 93\\
			\!\texttt{MI Surv}\!&						&										&										&													&				80								&								 80	&				80								&														& 80\\
			\!\texttt{MI Recur}\!&					&										&										&													&													&								 		&				80								&														& 80\\
			\!\texttt{UA}\!&								&										&										&													&													&								 		&													&														&   \\
			\!\texttt{Early HF}\!&					&										&										&													&													&								 		&													&		80											& 80\\
			\!\texttt{Severe HF}\!&					&										&										&													&													&								 		&													&														& 80\\
			\bottomrule	
			\end{tabular}
\end{table}

\begin{flushleft}
\begin{table}
\caption{\label{table:RR_Aspirin} Risk reductions for aspirin (\%)}
		\begin{tabular}{|llllllllll|}
			\toprule
								&	\!\texttt{AMI}\!\!	& \!\texttt{SD}\!\!	& \!\texttt{CA}\!\!	& \!\texttt{MI Surv}\!\!	& \!\texttt{MI Recur}\!\!	& \!\texttt{UA}\!\!	& \!\texttt{Early HF}\!\!	& \!\texttt{Severe HF}\!\!	& \!\texttt{CHD Death}\!\!\\
			\midrule
	 		\!\texttt{AMI}\!&								&										&										&						85						&													&										&				85								&														& 85\\
			\!\texttt{SD}\!&		85					&										&										&													&													&								 		&													&														& 85\\
			\!\texttt{CA}\!&		85					&										&										&													&													&								 85	&				85								&														& 85\\
			\!\texttt{MI Surv}\!&						&										&										&													&				85								&								 85	&				85								&														& 85\\
			\!\texttt{MI Recur}\!&					&										&										&													&													&								 		&				85								&														& 85\\
			\!\texttt{UA}\!&		85					&										&										&						93						&													&								 		&				85								&														&   \\
			\!\texttt{Early HF}\!&					&										&										&													&													&								 		&													&		85											& 85\\
			\!\texttt{Severe HF}\!&					&										&										&													&													&								 		&													&														& 85\\
			\bottomrule
			\end{tabular}
\end{table}
\end{flushleft}

\begin{flushleft}
\begin{table}
\caption{\label{table:RR_BetaBlockers} Risk reductions for beta blockers (\%)}
		\begin{tabular}{|llllllllll|}
			\toprule
								&	\!\texttt{AMI}\!\!	& \!\texttt{SD}\!\!	& \!\texttt{CA}\!\!	& \!\texttt{MI Surv}\!\!	& \!\texttt{MI Recur}\!\!	& \!\texttt{UA}\!\!	& \!\texttt{Early HF}\!\!	& \!\texttt{Severe HF}\!\!	& \!\texttt{CHD Death}\!\!\\
			\midrule
	 		\!\texttt{AMI}\!&								&										&										&						96						&													&										&				96								&														& 96\\
			\!\texttt{SD}\!&								&										&										&													&													&								 		&													&														&   \\
			\!\texttt{CA}\!&		77					&										&										&													&													&								 77	&				77								&														& 77\\
			\!\texttt{MI Surv}\!&						&										&										&													&				77								&								 77	&				77								&														& 77\\
			\!\texttt{MI Recur}\!&					&										&										&													&													&								 		&				77								&														& 80\\
			\!\texttt{UA}\!&								&										&										&													&													&								 		&													&														&   \\
			\!\texttt{Early HF}\!&					&										&										&													&													&								 		&													&		65											& 65\\
			\!\texttt{Severe HF}\!&					&										&										&													&													&								 		&													&														& 65\\
			\bottomrule	
			\end{tabular}
\end{table}
\end{flushleft}

\begin{flushleft}
\begin{table}
\caption{\label{table:RR_Statins} Risk reductions for statins (\%)}
		\begin{tabular}{|llllllllll|}
			\toprule
							&	\!\texttt{AMI}\!\!	& \!\texttt{SD}\!\!	& \!\texttt{CA}\!\!	& \!\texttt{MI Surv}\!\!	& \!\texttt{MI Recur}\!\!	& \!\texttt{UA}\!\!	& \!\texttt{Early HF}\!\!	& \!\texttt{Severe HF}\!\!	& \!\texttt{CHD Death}\!\!\\
			\midrule
	 		\!\texttt{AMI}\!&								&										&										&													&													&										&													&														&   \\
			\!\texttt{SD}\!&								&										&										&													&													&								 		&													&														&   \\
			\!\texttt{CA}\!&		78					&										&										&													&													&								 78	&													&														&   \\
			\!\texttt{MI Surv}\!&						&										&										&													&				78								&								 78	&				78								&														& 78\\
			\!\texttt{MI Recur}\!&					&										&										&													&													&								 		&				78								&														& 78\\
			\!\texttt{UA}\!&								&										&										&													&													&								 		&													&														&   \\
			\!\texttt{Early HF}\!&					&										&										&													&													&								 		&													&		78											& 78\\
			\!\texttt{Severe HF}\!&					&										&										&													&													&								 		&													&														& 78\\
			\bottomrule	
			\end{tabular}
\end{table}
\end{flushleft}

\subsection{Results}
Recall that the model presented in this paper will simulate a given number of individuals until death. That is, a record is produced of the path that each individual takes through the network with a list of states visited and times of transition. This information can be manipulated in numerous different way, some of which are shown here. The type of information and the format in which it is presented will depend on the particular user. For example, plots of certain output statistics may be more useful for model development in order to validate and diagnose the model, whereas other representation may be more useful to policy makers with specific questions and (usually) little statistical knowledge.

\begin{figure}
	\centering
		\includegraphics[width=10cm]{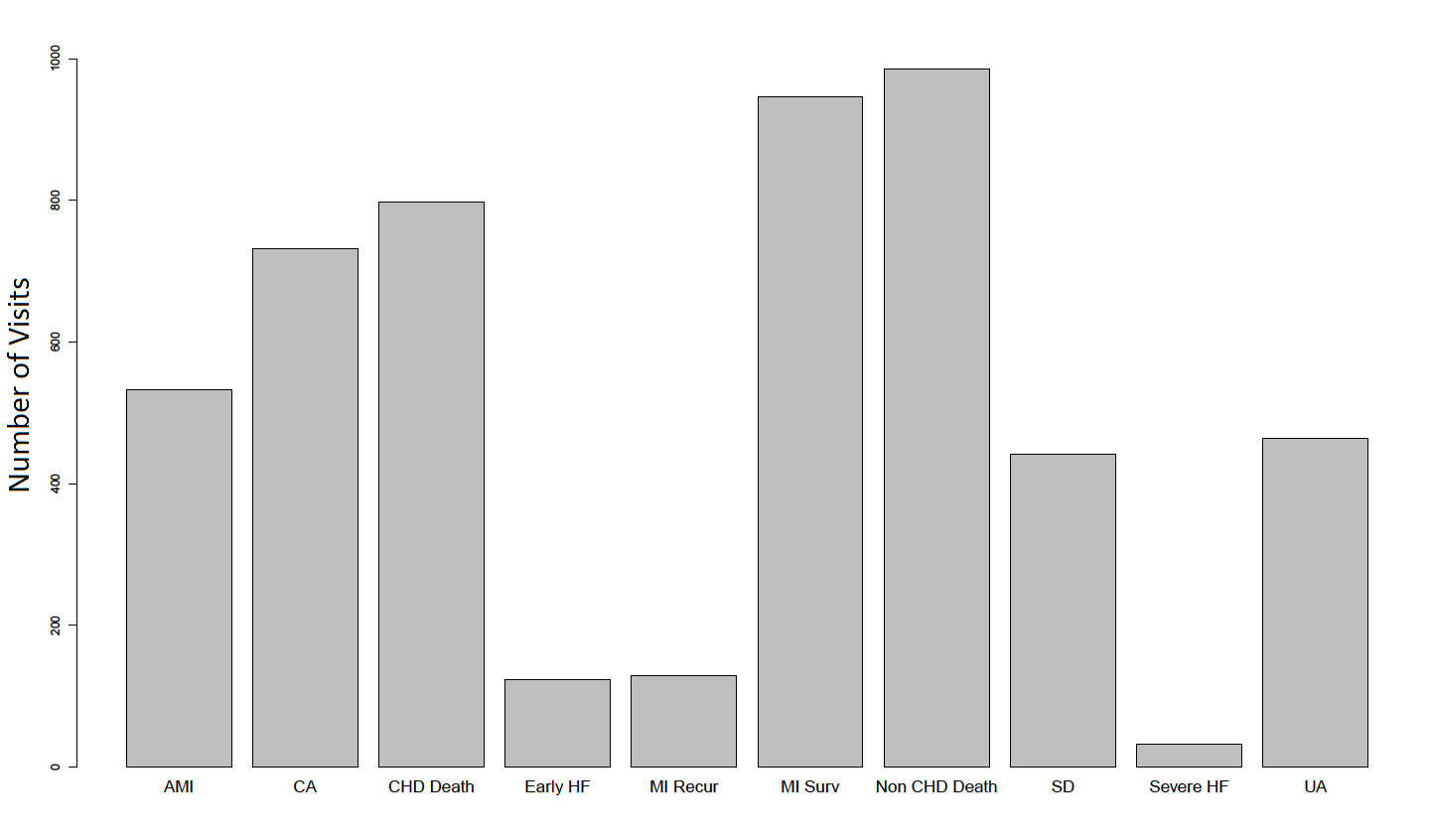}
				\caption{A bar chart showing the number of visits paid to each state during the course of the scenario when statins, beta blockers, aspirin and ACE inhibitors are all available.}
	\label{fig:numvisits_all}
\end{figure}

\begin{figure}
	\centering
		\includegraphics[width=10cm]{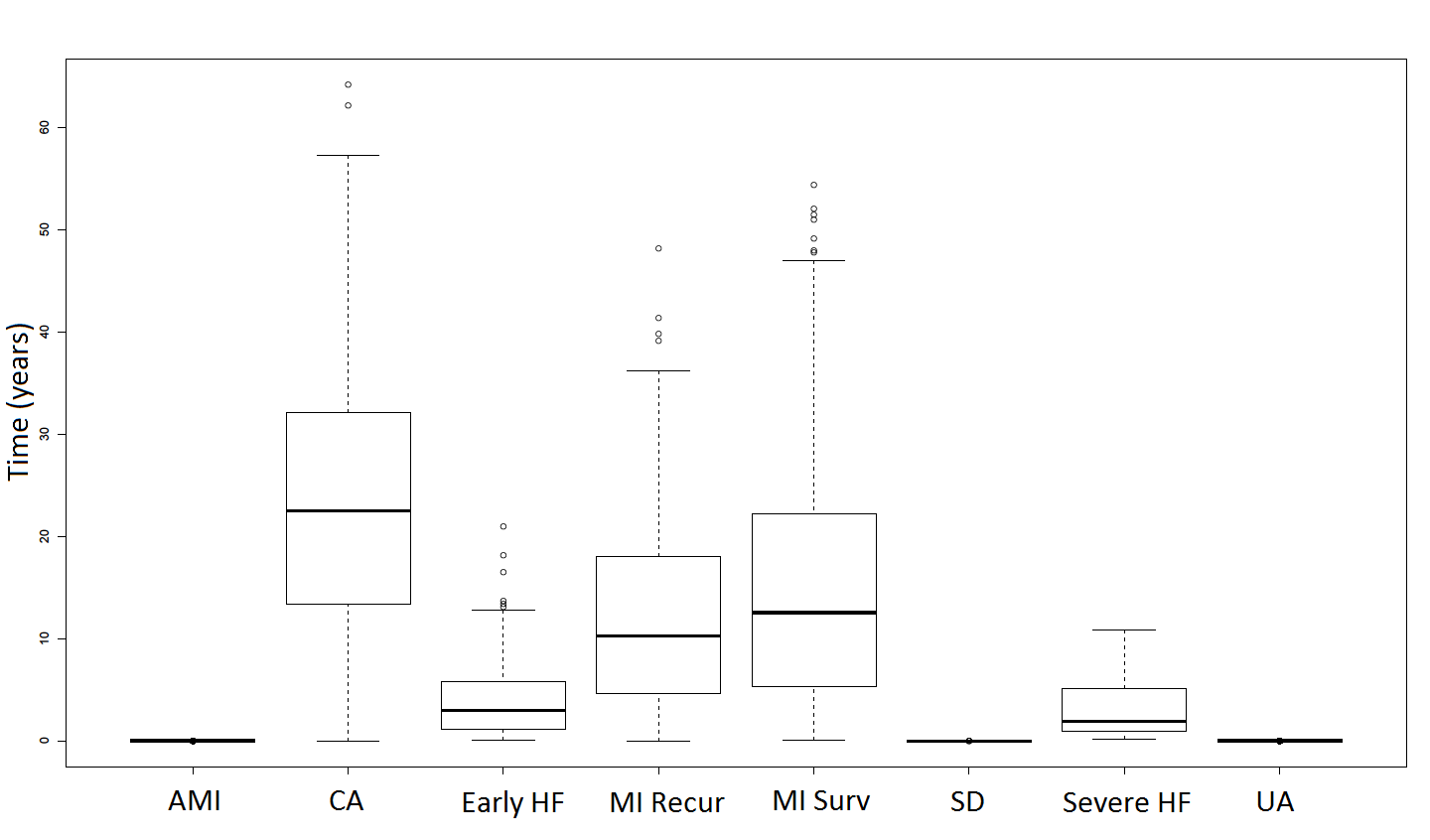}
				\caption{A box and whisker plot showing the length of stay in state per visit that an individual takes in each state when statins, beta blockers, aspirin and ACE inhibitors are all available.}
	\label{fig:staylength_all}
\end{figure}

\begin{figure}
	\centering
		\includegraphics[width=10cm]{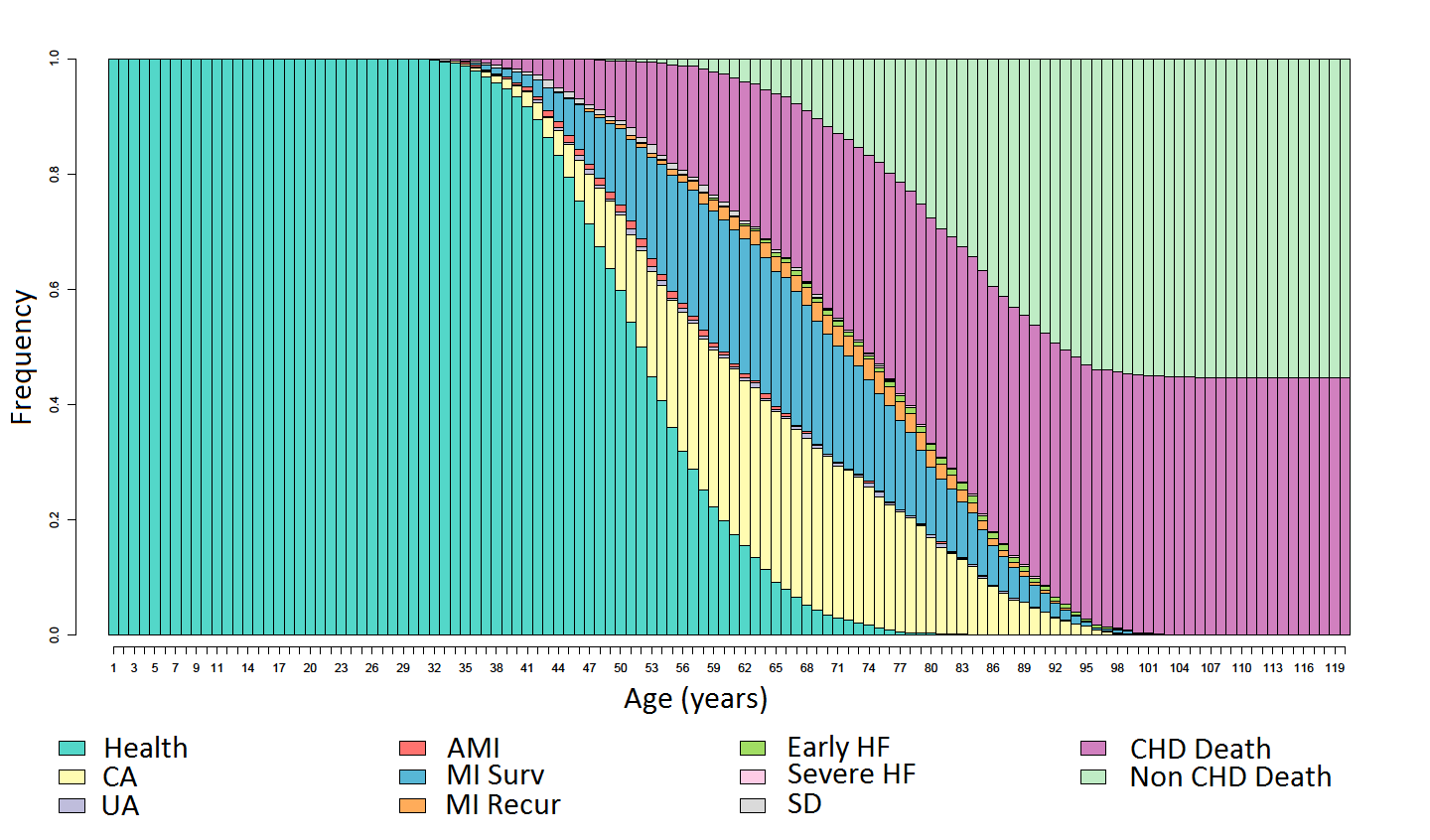}
				\caption{State volume against age when statins, beta blockers, aspirin and ACE inhibitors are all available.}
				\label{fig:statevol_all}
\end{figure}

\begin{figure}
	\centering
		\includegraphics[width=10cm]{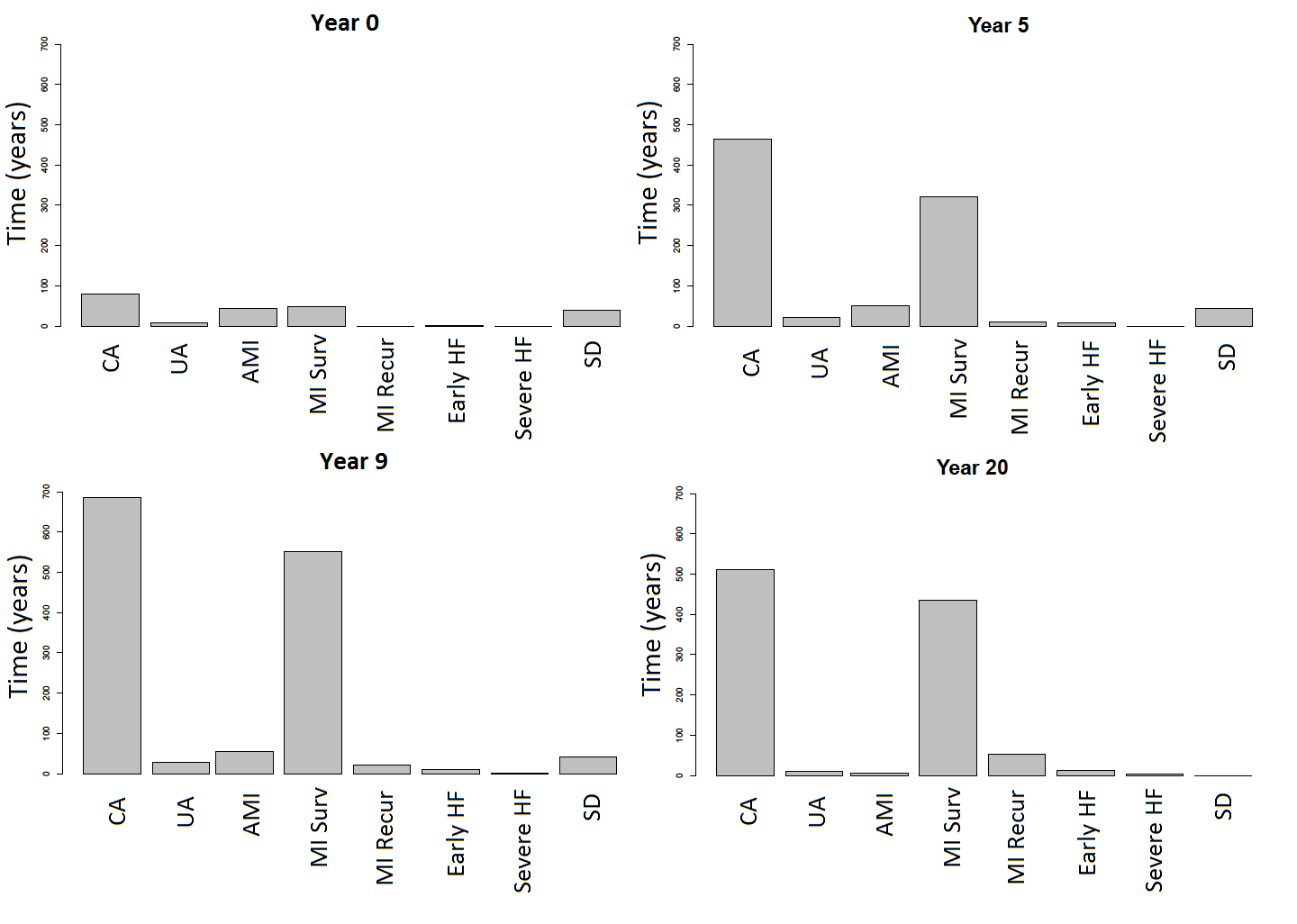}
				\caption{Prevalence when statins, beta blockers, aspirin and ACE inhibitors are all available.}
				\label{fig:prevalence_all}
\end{figure}

\begin{figure}
	\centering
		\includegraphics[width=10cm]{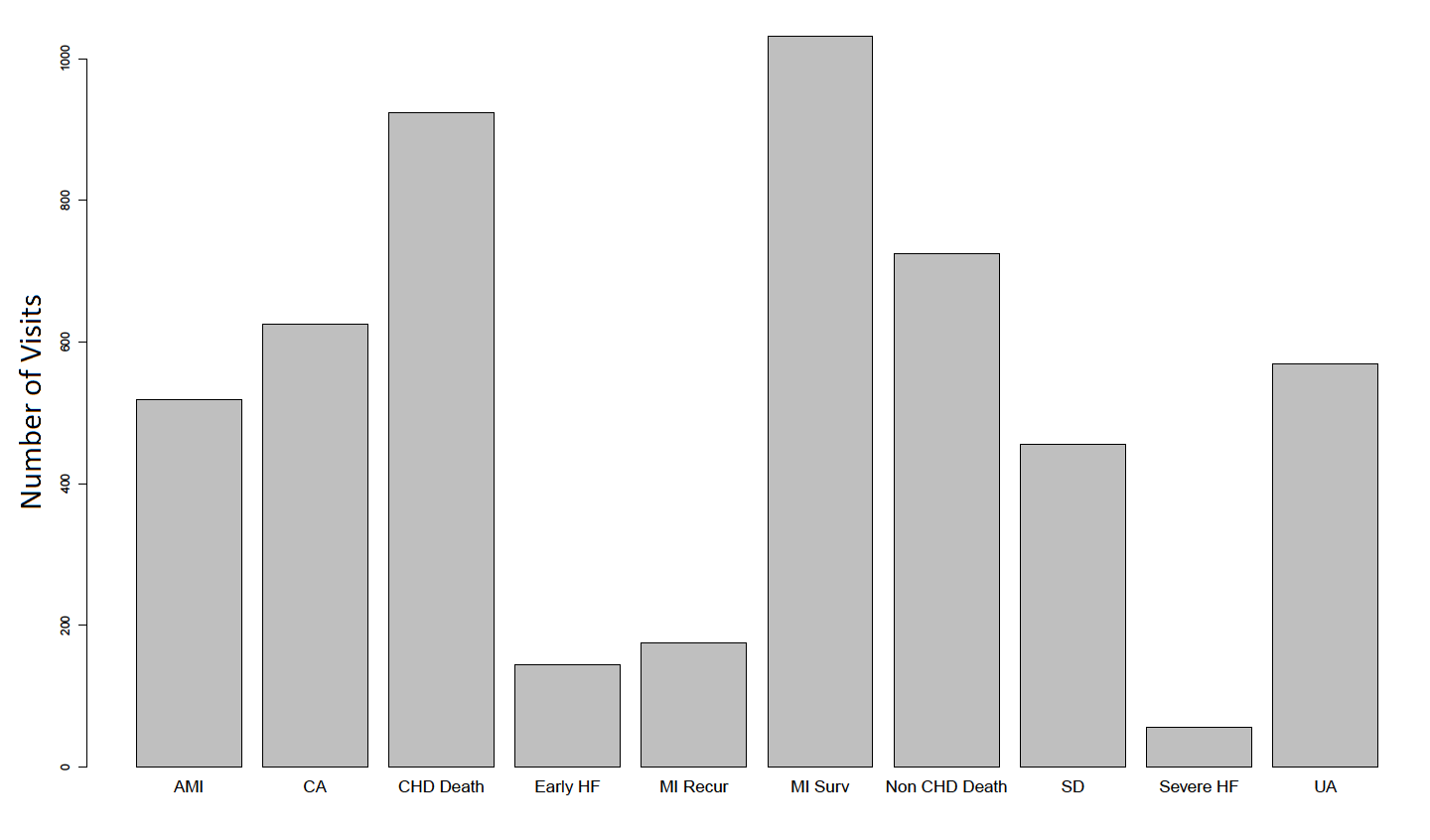}
				\caption{A bar chart showing the number of visits paid to each state during the course of the scenario when only beta blockers are available.}
	\label{fig:numvisits_BB_only}
\end{figure}

\begin{figure}
	\centering
		\includegraphics[width=10cm]{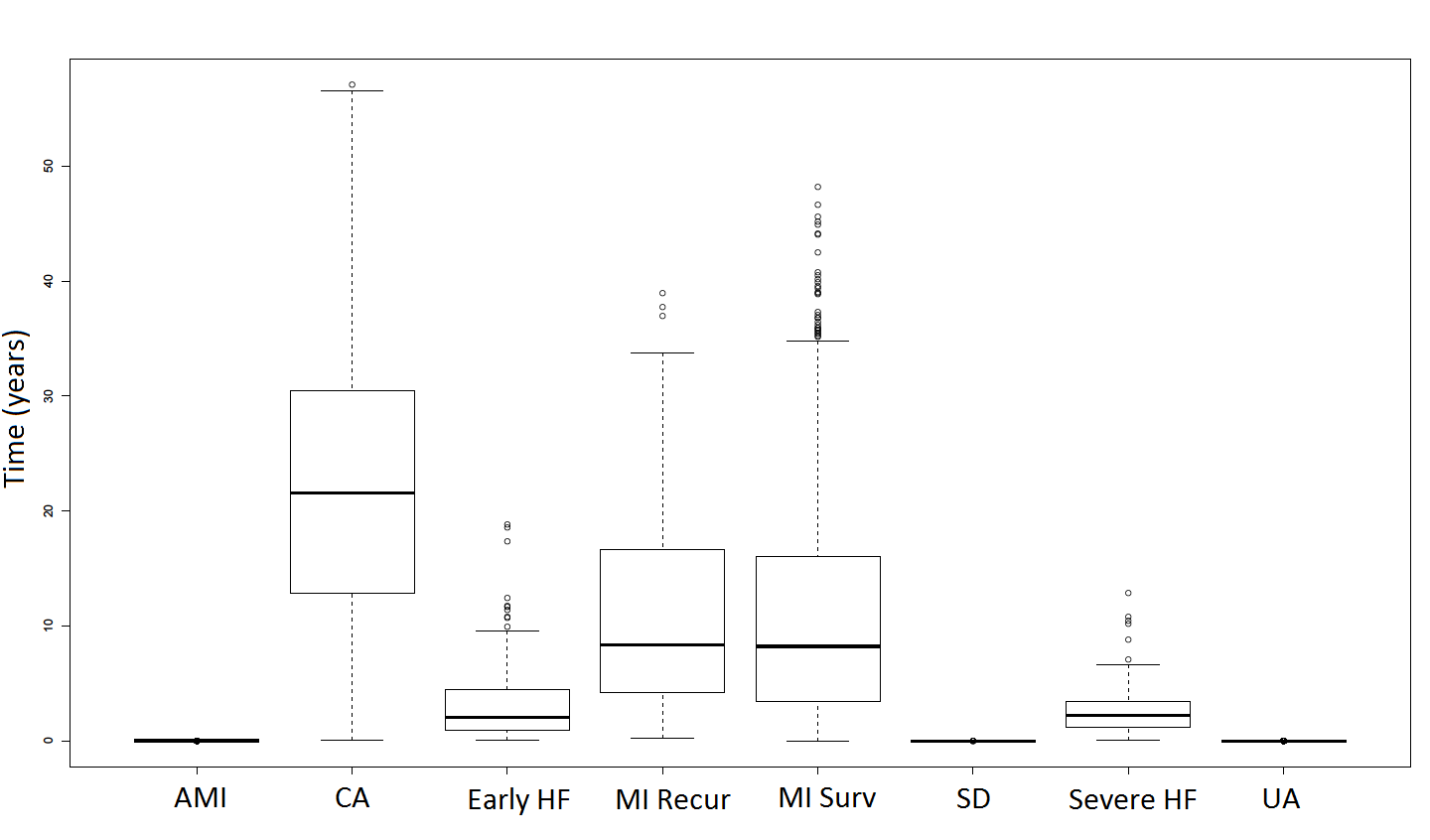}
				\caption{A box and whisker plot showing the length of stay in state per visit that an individual takes in each state when only beta blockers are available.}
	\label{fig:staylength_BB_only}
\end{figure}

\begin{figure}
	\centering
		\includegraphics[width=10cm]{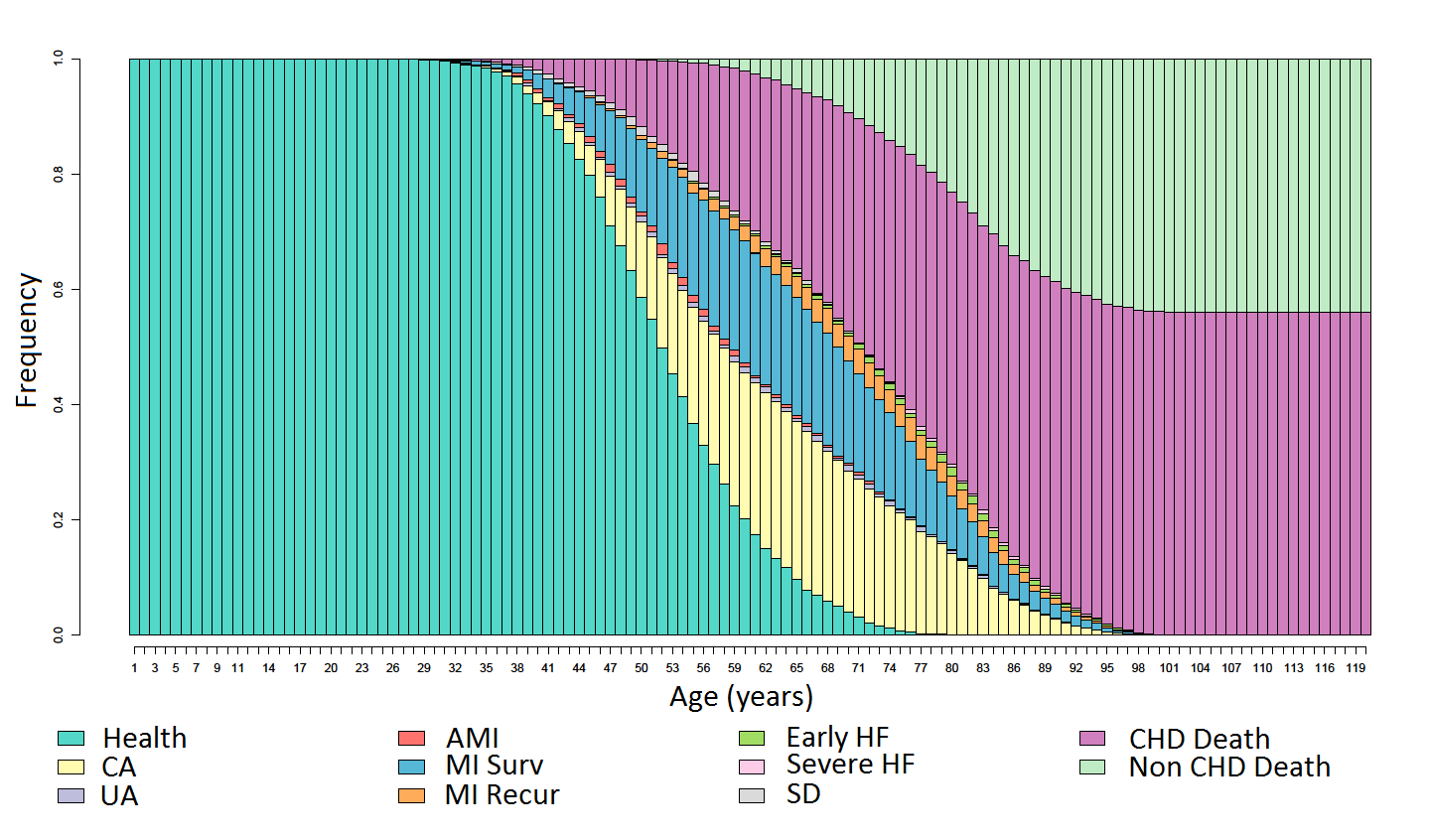}
				\caption{State volume against age when only beta blockers are available.}
				\label{fig:statevol_BB_only}
\end{figure}

\begin{figure}
	\centering
		\includegraphics[width=10cm]{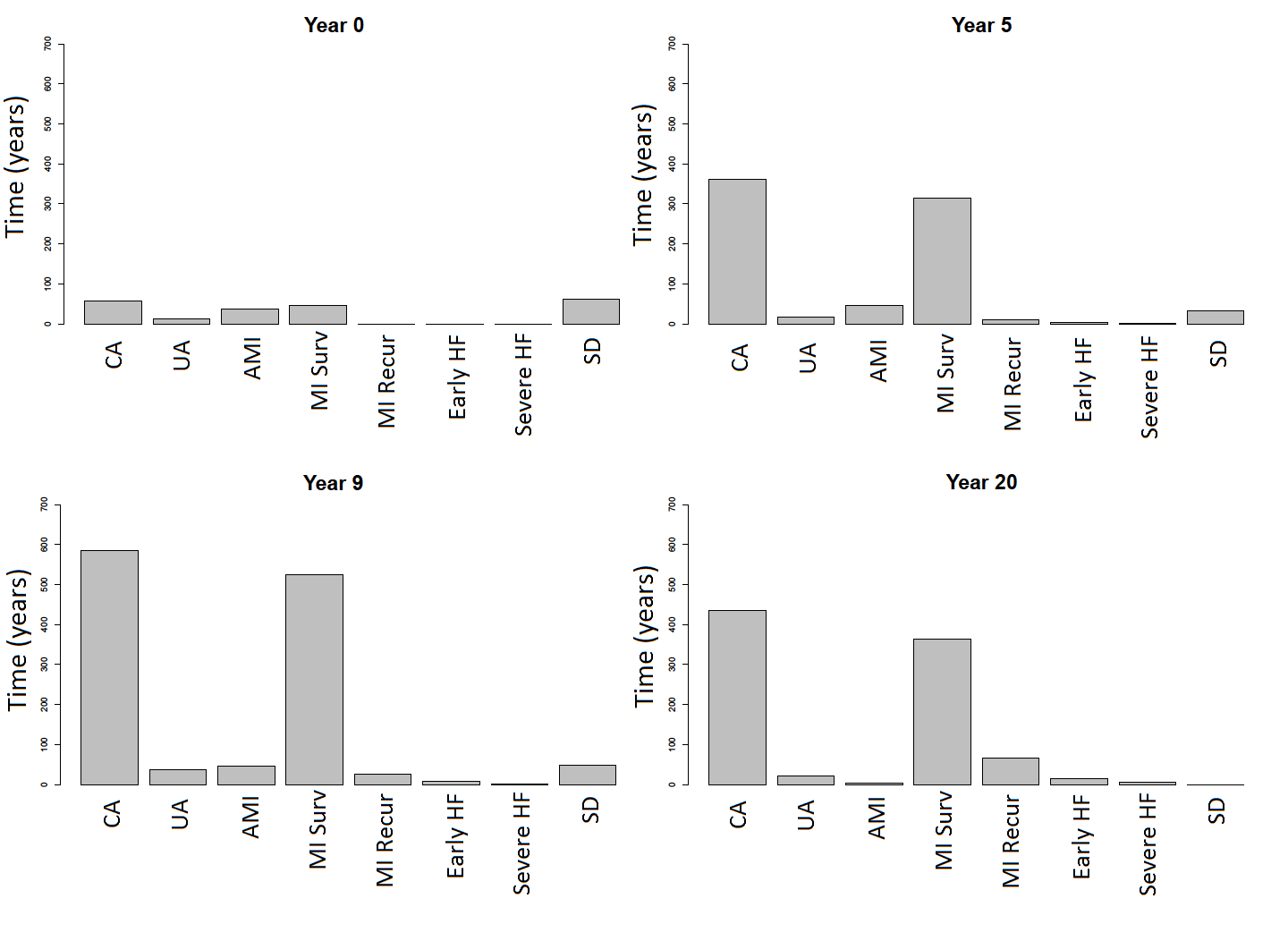}
				\caption{Prevalence when only beta blockers are available.}
				\label{fig:prevalence_BB_only}
\end{figure}

\begin{figure}
	\centering
		\includegraphics[width=10cm]{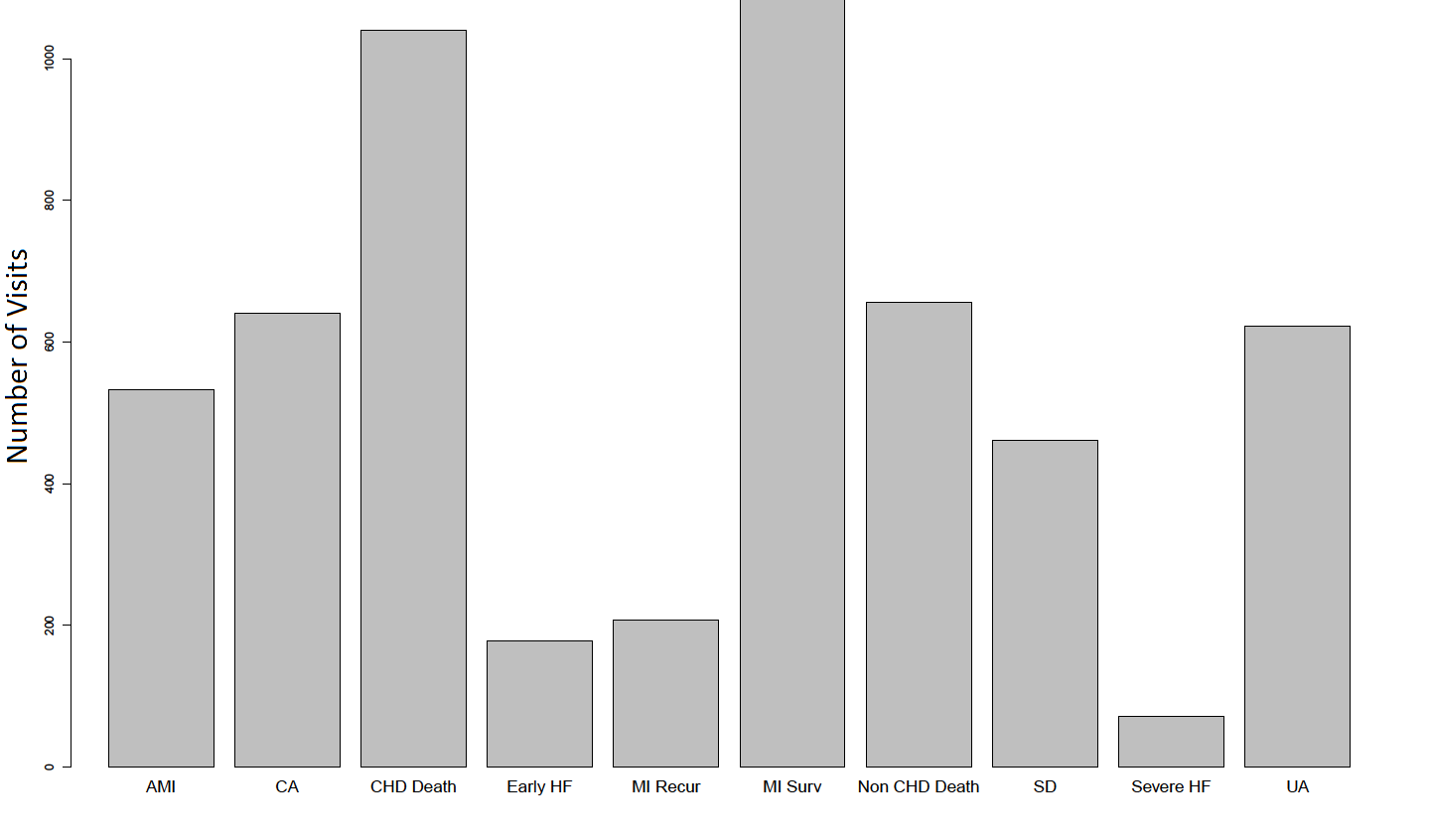}
				\caption{A bar chart showing the number of visits paid to each state during the course of the scenario with no interventions.}
	\label{fig:numvisits_Baseline}
\end{figure}

\begin{figure}
	\centering
		\includegraphics[width=10cm]{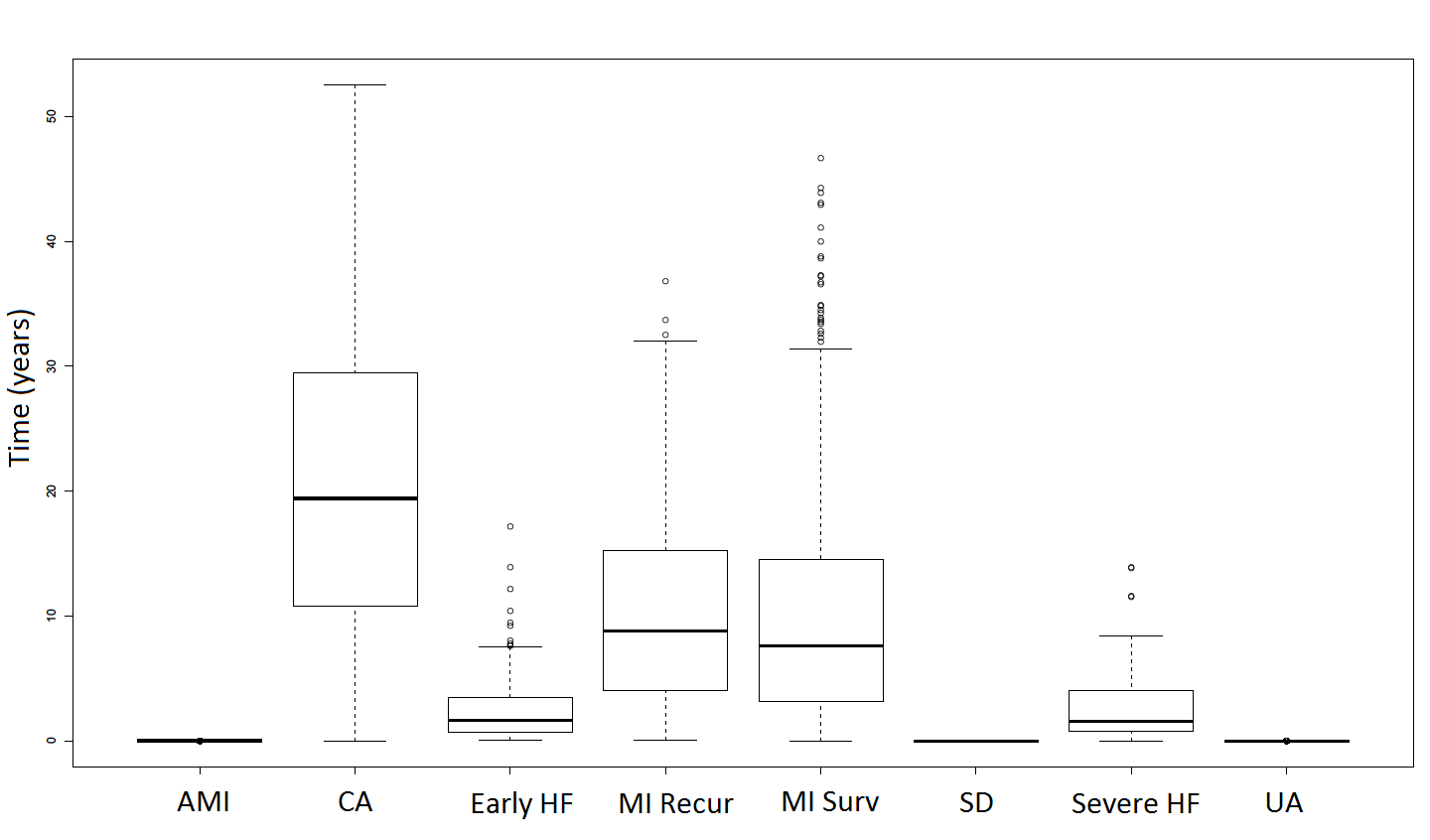}
				\caption{A box and whisker plot showing the length of stay in state per visit that an individual takes in each state for the scenario with no interventions.}
	\label{fig:staylength_Baseline}
\end{figure}

\begin{figure}
	\centering
		\includegraphics[width=10cm]{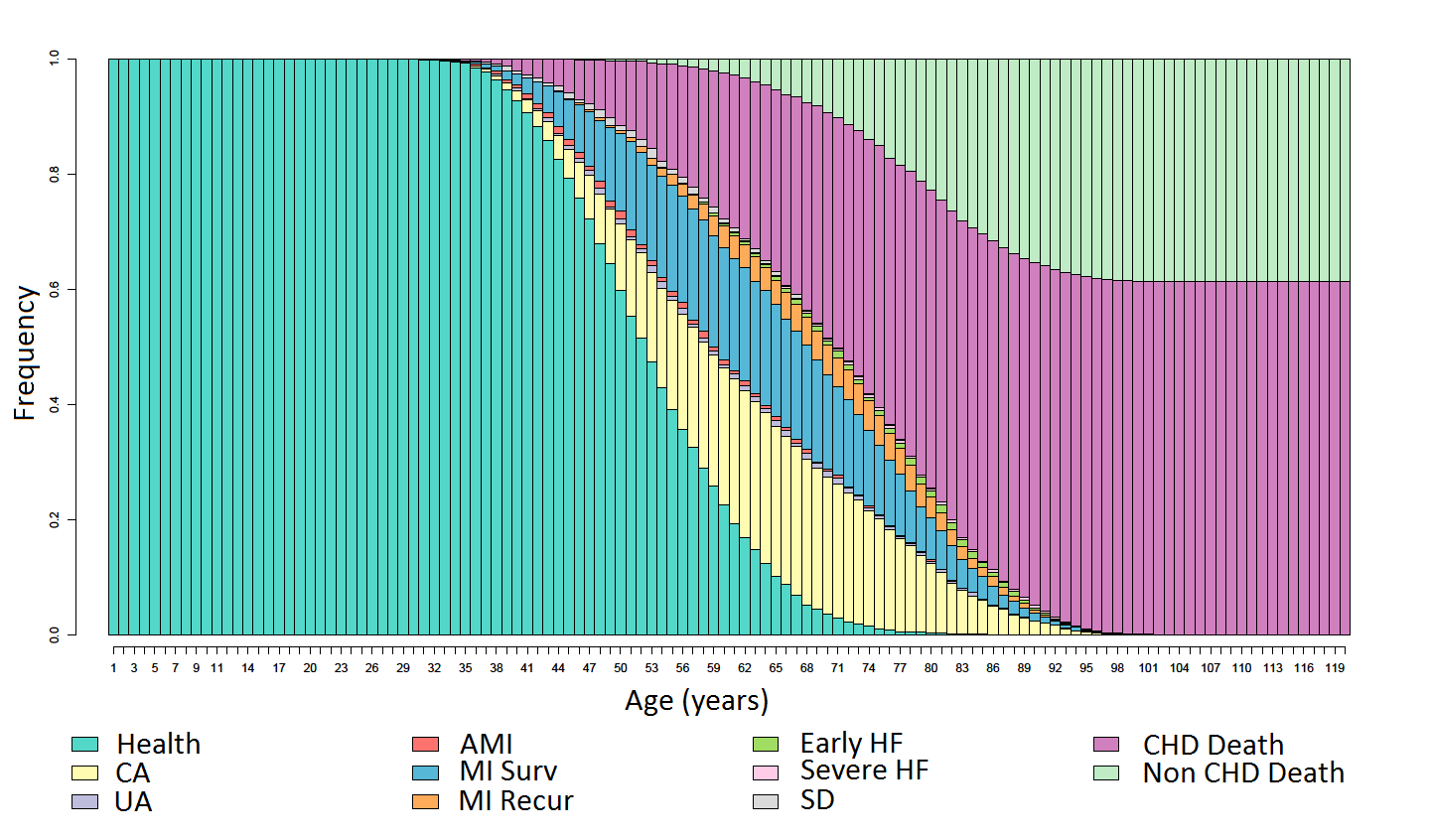}
				\caption{State volume against age for the scenario with no interventions.}
				\label{fig:statevol_Baseline}
\end{figure}

\begin{figure}
	\centering
		\includegraphics[width=10cm]{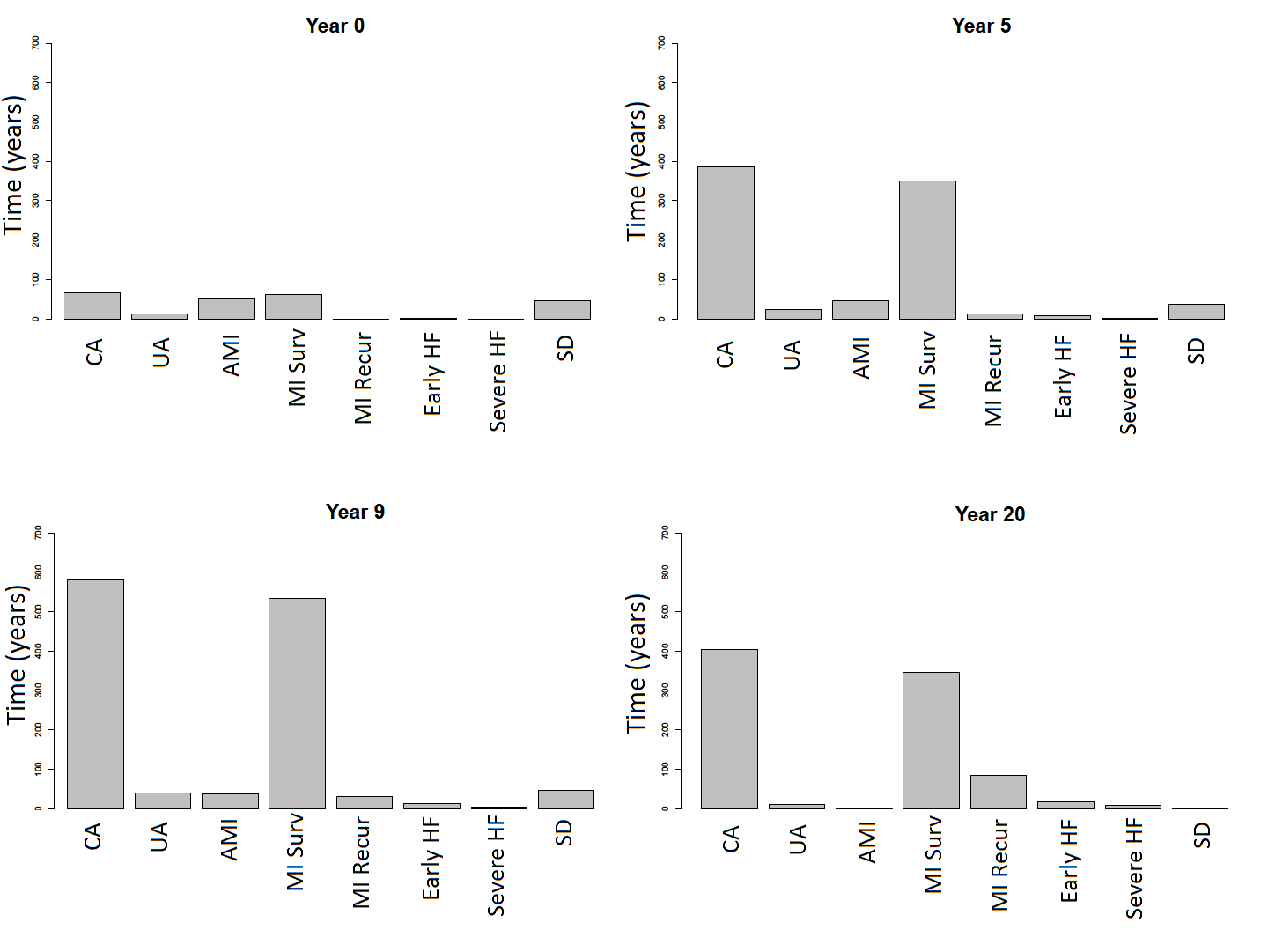}
				\caption{State prevalence for the scenario with no interventions.}
				\label{fig:prevalence_Baseline}
\end{figure}

Figures~\ref{fig:numvisits_all}, \ref{fig:staylength_all}, \ref{fig:statevol_all} and \ref{fig:prevalence_all} give possible output presentation for the case where statins, beta blockers, aspirin and ACE inhibitors are all available.
Figures~\ref{fig:numvisits_BB_only}, \ref{fig:staylength_BB_only}, \ref{fig:statevol_BB_only} and \ref{fig:prevalence_BB_only} give some output presentation for the case where only beta blockers are available.
Figures~\ref{fig:numvisits_Baseline}, \ref{fig:staylength_Baseline}, \ref{fig:statevol_Baseline} and \ref{fig:prevalence_Baseline} give possible output presentation for the case where no interventions are available.  

Using a Cox regression approach \cite{Collett2003}, Table~\ref{table:coxregression} gives relative risk reductions for the intervention strategies compared against the baseline simulation, where the outcome of interest is time to CHD death. Relative risk reductions are given, with their standard errors in braces. The $p$-values correspond to the null hypothesis that the risk reductions are not different from zero.

\begin{table}
\caption{\label{table:coxregression} Relative risk reductions for the intervention strategies compared against the baseline simulation}
	\begin{tabular}{|lll|}
	\toprule
	Intervention & Relative Risk Reduction & $p$-value\\
	\midrule
	Baseline &0 &\\
	ACEI &0.101 (0.040) &0.015\\
	Aspirin &0.016 (0.044) &0.710\\
	Beta Blockers &0.137 (0.039) &0.001\\
	Statins &0.115 (0.039) &0.006\\
	ALL &0.356 (0.030) &0.000\\
	\bottomrule
	\end{tabular}
\end{table}
If the interventions were acting independently, a Mant-Hicks calculation \cite{Mant_Hicks1995} of the relative risk reduction expected for the combination of all interventions would be $1-\big\{(1-0.101)\times(1-0.016)\times(1-0.115)\big\} = 0.324$. This is well within a $95\%$ confidence interval of the value for all interventions --- $(0.297,0.415)$.

\section{Discussion}
\label{section:discussion}
Figures~\ref{fig:numvisits_all}, \ref{fig:numvisits_BB_only} and \ref{fig:numvisits_Baseline} show the period prevalence for the full intervention, beta blockers only and no intervention scenario respectively.
It can be seen that the effect of more interventions are to decrease the number of CHD related deaths and consequently increase the number of non-CHD related deaths. Also, with the increased amount of preventative treatment, the occurrence of unstable angina is reduced. 

Figures~\ref{fig:staylength_all}, \ref{fig:staylength_BB_only} and \ref{fig:staylength_Baseline} show the sojourn times at each state visit for the full intervention, beta blockers only and no intervention scenario respectively. We can see that the length of time in MI Survival increases with more treatment. This is because patients are less likely to enter an acute state.
Furthermore, the interquartile range for the length of time in state for Early and Severe Heart Failure and, in particular, MI Survival is greater and with fewer outliers for the treatment scenarios. This is due to the patients under-going treatment remaining in the model for longer periods, i.e. reaching an older age, but then they face a steep increase in non-CHD death risk. This effect is known as shouldering.

Figures~\ref{fig:statevol_all}, \ref{fig:statevol_BB_only} and \ref{fig:statevol_Baseline} show the state volume against age for the full intervention, beta blockers only and no intervention scenario respectively.
Again, we can see the increase in the number of non-CHD deaths in the scenarios with more treatments. Note that the Healthy region is identical between plots since this is the entry state of the model and so not affected by secondary interventions.

Figures~\ref{fig:prevalence_all}, \ref{fig:prevalence_BB_only} and \ref{fig:prevalence_Baseline} show the period prevalence against age for the full intervention, beta blockers only and no intervention scenario respectively.
The plots show that the period prevalence between scenarios is quite similar. Is seems that many of the paths followed by the patients are similar between scenarios. However, there appear to be more individuals in the chronic angina state for the full treatment scenario, possibly due to these patients not entering a more serious state.

\section{Conclusions}
\label{section:conclusions}
The design of a chronic disease policy model for prognostic reasoning (making a prediction about what will happen in the future) \cite{BioBayesNets} has been detailed.
In particular, a CHD model example has been presented and example output discussed.

The main features are the following:
\begin{itemize}
	\item The model is a patient care model which simulates the patient trajectory from the onset of the disease to death.
	\item Interventions strategies can be explored both at a population level and an individual level.
	\item A novel model fitting methodology has been developed to specify a sufficiently accurate model using historical data and expert judgment. 
\end{itemize}
In particular, the strengths of this model are:
\begin{itemize}
	\item The model is dynamic in that it represents the patient journey through time. Patient histories may be utilised to adjust transition probabilities in the future.
	\item The specification of the model using `constraints' is very flexible. This allows for the inclusion of prior knowledge in a simple form which is intuitive for practitioners.
	\item The use of (local) hazard functions admits a powerful and flexible model structure to the whole model.
	\item The graphical representation of the discrete event simulator is intuitive for clinicians and policy makers, thereby making it easier for them to develop models and interpret results.  
\end{itemize}
In order to build the model certain assumptions have been necessary. These are listed below.
\begin{itemize}
	\item \emph{Focussed context}: One important point is that, because we only consider CHD patients, we can only consider the public health burden in isolation, and not the overall public health burden. This means that caution is needed in interpreting the results of interventions.

	\item \emph{Constraint specification}: In the generation of the constraints, there is an assumption made about the size of the age interval. It is thus assumed that individuals share sufficiently similar characteristics within these bands to justify pooling them together.

	\item \emph{Model assumptions}: The patient life-courses and the adherences to interventions are assumed to be independent.
\end{itemize}

\subsection{Future work}
The modelling approach presented in this paper is an effectively structured synthesis of evidence in the absence of relevant individual-level health data. The above method using specified constraints is just one way of translating the data into a model parameter estimate. If we had access to other sources of data, it could be incorporated directly or used to estimate some other quantity of interest for the process, which would mean a different fitness function would be needed (see Section~\ref{section:fitness}).

Note that the constraints have been produced from data but there is no indication currently of the uncertainties on these estimates, such as upper and lower bounds. In a standard Bayesian approach, we would place prior distributions on the parameters determining the static transition probabilities. Then these would be updated in light of the data. Alternatively, we could place priors on the parameters of the hazard mixture and then update these after seeing the new data. However, the way to specify such prior distributions is a significant challenge because of the unintuitive representation for clinicians and decision makers.

Several model extensions could be made in terms of how interventions are applied. Interventions need not be singular, static events but may change during the time course of the scenario. It is more realistic to consider the application and cessation of interventions at any time. This could be after some set amount of time in a state, at different age groups within a state or at particular years.

Further, adherence to a given intervention may be generalised to consider more complex temporal patterns. For example, interventions may be halted at times depending on length of stay in state, age group or particular age, and alternative interventions may be introduced at this time.

The assumption of adherence independence may also be relaxed. For example, it is clearly more likely that an individual who has had an intervention for a long period is more likely to continue with another intervention.

Finally, in practice, limitation of funds will restrict the availability of particular interventions. Health policy scenarios may even be considered in terms of adding or removing a specific amount of funding from one intervention option to another. So the outputs of this model should be presented in economic as as well as health terms.

\section{Acknowledgments}
This work was carried out under the Greater Manchester Collaboration for Leadership and Applied Health Research and Care (CLAHRC) project, funded jointly by the National Institute for Health Research (NIHR) and the National Health Service (NHS).

We thank Simon Capewell, Martin O'Flaherty of the University of Liverpool for their support and public health expertise. We also thank the software development team of John Ainsworth, Richard Williams, Philip Couch and Emma Carruthers.

\bibliographystyle{plain}

\newpage
\appendix

\section{Model step-by-step algorithm description}
\label{chapter:stepbystep_algorithm}

\begin{table}[ht!]
\begin{flushleft}
		\caption{Graph setup}
		\begin{tabular}{ll}
			\hline \hline
			Step & Description \\
			[0.5ex]
			\hline
			1 & Define graph $(V,E)$ (nodes $v \in V$ and edges $e \in E$).\\
			2 & Specify model family $\mathcal{M}$.\\
			3 & Specify $N$ constraints, $\tilde{\phi}^j(t_j,\Delta_j), j=1,2, \ldots, N$.\\
			4	& Fit model by selecting the optimal model,\\
				& $M^* = \stackrel[M \in \mathcal{M}]{}{\textrm{argmin}} \sum_{j=1}^N w_j d \left( \tilde{\phi}^j(t_j,\Delta_j), \widehat{\phi}^j(t_j,\Delta_j)(M) \right)$.
			\end{tabular}
\end{flushleft}
\end{table}

\begin{table}[ht!]
\begin{flushleft}
		\caption{Simulation without interventions}
		\begin{tabular}{ll}
			\hline \hline
			Step & Description \\
			[0.5ex]
			\hline
			5  & Set maximum scenario duration $t_{max}$.\\	
			6  & a) Define initial number of cases $S$, depending on\\
				 & the total population size and the duration of the simulation.\\
				 &  b) Define the data for each individual $s$:\\
				 & gender, age, start time $t^s_0$ and state $x_0^s$, either from the primary side or a closed cohort.\\ 
			7  & Set case count $s=1$.\\
			8  & Set transition count $i=0$.\\ 
			9  & Generate a transition type $C$ and time $T$ from the competing risks, $R_{x_i^s}$,\\
				 & using the sub-hazard functions,\\
				 & where $T = \min_r\{ T_r ; r=1,2,\ldots,R_{x_i^s} \}$ and $C=\mbox{argmin}_r\{T_r;r=1,2,\ldots,R_{x_i^s} \}$.\\
			10 & Set $t^s_{i+1}=T$ and $x_{i+1}^s=C$.\\
			11 & If $t^s_{i+1} > t_{max}$ and $s<S$ then let $s=s+1$ and go to Step 8.\\	
			12 & If $x_{i+1}^s \in \{ \texttt{CHD Death}, \texttt{Non CHD Death} \}$ and $s<S$ then let $s=s+1$ and go to Step 8.\\
			13 & Set $i=i+1$ and go to Step 9.
			\end{tabular}
\end{flushleft}
\end{table}

\begin{table}[ht!]
\begin{flushleft}
		\caption{Simulation with interventions}
		\begin{tabular}{ll}
			\hline \hline
			Step & Description \\
			[0.5ex]
			\hline
			14 & Specify $L$ intervention types $\omega_1, \omega_2, \ldots, \omega_L$.\\ 
			15 & For each gender $\lambda \in \{M, F\}$ and node $v \in V$,\\
				 & define a set of interventions $\bm{\tau}^{\lambda,v} \subseteq \{ \omega_l :  l=1,\ldots,L \}$.\\ 
			16 & Specify the probability of adherence depending on intervention, gender and state $\nu_{\omega_l}^{\lambda,v}$.\\
			17 & Define relative risk reductions on each edge $e \in E$, associated with each $\tau^{\lambda,v}$ by $\bm{\eta}^{\lambda,v}_{\tau}$. \\ 
			18 & Set maximum scenario duration $t_{max}$.\\	 
			19 & a) Define initial number of cases $S$, depending on\\
				 & the total population size and the duration of the simulation.\\
				 &  b) Define the data for each individual $s$:\\
				 & gender, age, start time $t^s_0$ and state $x_0^s$,\\
				 & either generated from the pre-clinical model or a closed cohort.\\ 
			20 & Set case count $s=1$ and interventions applied to $s$ by the empty set, $\Omega_s = \phi$.\\ 
			21 & Set transition count $i=0$.\\ 
			22 & For each intervention $\tau^{\lambda^s, x_i^s}$ then, \\
				 & $\Omega_s = \{ \Omega_s, \bm{\eta}^{\lambda^s, x_i^s}_{\tau} \}$ with probability $\nu_{\tau}^{\lambda^s, x_i^s}$.\\
			23 & Generate a transition type $C$ and time $T$ from the competing risks, $R_{x_i^s}$,\\
				 & using the (proportionally adjusted) sub-hazard functions $h^{\Omega_s}_r(t), r=1,2,\ldots,R_{x_i^s}$,\\
				 & where $T = \min_r\{ T_r ; r=1,2,\ldots,R_{x_i^s} \}$ and $C=\mbox{argmin}_r\{T_r;r=1,2,\ldots,R_{x_i^s} \}$,\\
			24 & Set $t^s_{i+1}=T$ and $x_{i+1}^s=C$.\\
			25 & If $t^s_{i+1} > t_{max}$ and $s<S$ then let $s=s+1$ and go to Step 22.\\	
			26 & If $x_{i+1}^s \in \{ \texttt{CHD Death}, \texttt{Non CHD Death} \}$ and $s<S$ then let $s=s+1$ and go to Step 22.\\
			27 & Set $i=i+1$ and go to Step 23.		
			\end{tabular}
\end{flushleft}
\end{table}

\end{document}